# Laser-cooling-assisted mass spectrometry

Christian Schneider,[1, *] Steven J. Schowalter,[1] Kuang Chen,[1] Scott T. Sullivan,[1] and Eric R. Hudson[1]

[1]*Department of Physics and Astronomy, University of California, Los Angeles, California 90095, USA*

Mass spectrometry is used in a wide range of scientific disciplines including proteomics, pharmaceutics, forensics, and fundamental physics and chemistry. Given this ubiquity, there is a worldwide effort to improve the efficiency and resolution of mass spectrometers. However, the performance of all techniques is ultimately limited by the initial phase-space distribution of the molecules being analyzed. Here, we dramatically reduce the width of this initial phase-space distribution by sympathetically cooling the input molecules with laser-cooled, co-trapped atomic ions, improving both the mass resolution and detection efficiency of a time-of-flight mass spectrometer by over an order of magnitude. Detailed molecular dynamics simulations verify the technique and aid with evaluating its effectiveness. Our technique appears to be applicable to other types of mass spectrometers.

## I. INTRODUCTION

Mass spectrometry (MS) is an integral tool in modern science, constituting a roughly $ 4 billion annual market [1, 2]. It is used to identify assays in hospitals and forensics laboratories [3, 4], characterize complex proteins [5], discover and produce pharmaceuticals [6], and enable important measurements in fundamental physics and chemistry [7]. Given this ubiquity, there is a worldwide effort to develop mass spectrometers with ever-improving performance [8–12].

The performance of a mass spectrometer is primarily characterized by two numbers: the mass resolution, calculated as $m/\Delta m$, where $m$ is the mass of the molecule being analyzed and $\Delta m$ is the imprecision of the mass measurement, and the detection limit, which is the minimum number of molecules required to produce a useful signal in the spectrometer. Fundamentally, a mass spectrometer operates by separating the initial phase-space volume of the molecules under study, i.e. the assay, into distinct phase-space volumes in a mass dependent way. For example, in a time-of-flight mass spectrometer (TOF-MS) mass dependent propagation delays are used to separate the initial phase-space volume, while in a Fourier-transform mass spectrometer the mass dependence of an orbital frequency is used. In an ideal mass spectrometer, each mass in the assay would be mapped to a single point in phase-space, allowing infinite mass resolution and detection of every molecule in the assay.

Unfortunately, the separation techniques employed in mass spectrometers are generally dissipationless and, thus, according to Liouville's theorem, the phase-space volume cannot be compressed. Therefore, the ideal mass-spectrometer performance can only be realized, if the molecular phase-space distribution begins with zero width. However, practical considerations for a mass spectrometer are at odds with this desire for a small initial phase space volume. Due to the relative ease of manipulating and detecting charged particles, modern mass spectrometers analyze molecules that have been vaporized and ionized. These ions are typically produced using techniques, such as electrospray ionization (ESI) [13, 14] and matrix-assisted laser desorption/ionization (MALDI) [15–17], which create ionized assays of input molecules with large kinetic energy, several $k_\text{B} \times (10^3\text{–}10^5)\,\text{K}$. To combat the large width of the phase-space distribution that comes with these elevated temperatures and improve performance, commercial mass spectrometers typically employ a collisional-cooling stage [12, 18–23] with an inert, room-temperature buffer gas, as originally demonstrated for guided ions in Ref. 24. This cooling narrows the input phase space distribution and improves, to an extent, mass spectrometer performance.

To further mitigate the effects of the initial phase-space distribution, a variety of techniques [8–12] has been developed, which effectively squeeze the phase space volume in one dimension and/or make the mass spectrometer insensitive to a spread in a certain dimension. Nonetheless, improved mass resolution using these techniques often comes at the price of a reduced detection efficiency (or vice versa), because practical limitations, such as machining imprecisions, ultimately limit their efficiency. Further, some techniques intentionally restrict mass analysis to only a fraction of the available phase-space distribution for increased resolution.

Here, we propose and demonstrate a new method, called laser-cooling-assisted mass spectrometry (LA-MS), to dramatically reduce the width of the molecular phase-space distribution by cooling the assay to temperatures up to six orders of magnitude lower than traditional room-temperature buffer-gas cooling. Input molecules are sympathetically cooled [25, 26] with laser-cooled, co-trapped atomic ions prior to mass analysis (hereafter referred to as LA-MS[X$^+$], where X$^+$ denotes the laser-cooled species). The spread of the phase-space distribution in momentum space is virtually eliminated, while the extent of the spatial component of the distribution is limited only by the balance of the confining potential holding the assay and the Coulomb repulsion between constituent ions. By implementing LA-MS in a traditional TOF-MS, we demonstrate an improvement in both mass resolution and detection limit of over an order of magnitude. A similar technique has been suggested to improve the mass spectrometry of highly-charged ions in a Penning trap by extending trap lifetimes and enabling longer interrogation times [27–29].

* christian.schneider@physics.ucla.edu

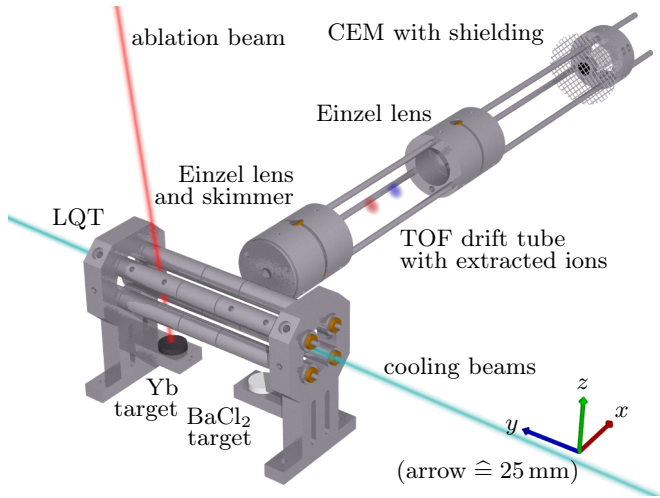

Figure 1. Schematic of the LA-MS apparatus. The LQT has a field radius of $R_0 = 6.85$ mm and a length of 91 mm. Its axis is aligned perpendicular to the TOF drift tube to enable the radial extraction. The Yb and BaCl$_2$ ablation targets are mounted below the LQT. The laser cooling beams are roughly aligned with the axis of the LQT. The TOF drift tube contains two Einzel lenses and has a total length of 275 mm. The CEM is shielded by a grounded stainless steel mesh and the complete assembly is held under vacuum at a pressure of $\approx 10^{-9}$ mbar. The fluorescence of the laser-cooled species is imaged through an objective lens (not shown), facing the TOF-MS, onto an electron-multiplying CCD camera and allows estimation of the number of atoms and molecules in the assay.

Although LA-MS is demonstrated with a TOF-MS, this technique appears to be applicable to other types of mass spectrometers. In its current form, the technique can be used to assay compounds up to a mass of approximately $10^3$ Da, which includes volatile organic compounds, the amino acids, explosive agents, and some peptides; modifications may extend the technique to include heavier biomolecules such as nucleic acids.

In the remainder of this article, we report on the results of a proof-of-principle experiment using LA-MS with a basic, Wiley-McLaren type TOF-MS [8]. Our TOF-MS does not utilize a reflectron [10] or other advanced engineering techniques so that the effectiveness of LA-MS can be judged independently of these techniques. Along with a discussion of the improvements in mass resolution, detailed molecular dynamics simulations are presented, which reproduce our experimental findings and aid evaluation of the effectiveness of LA-MS. We conclude with possible uses and future steps to be taken using this powerful new method.

## II. EXPERIMENTAL SETUP

The apparatus consists of a segmented linear quadrupole trap (LQT) and a TOF-MS, into which molecules can be radially extracted from the LQT (see Fig. 1; similar to the apparatus previously outlined in Ref. 30). The LQT has a field radius of $R_0 = 6.85$ mm and is driven asymmetrically (two diagonally opposing electrodes at RF voltage $V_0$, the others at RF ground) at a frequency of $\Omega \approx 2\pi \times 0.7$ MHz and an RF amplitude of $V_0 = 250$ V.

The LA-MS experimental sequence is as follows. Samples of atoms and molecules are loaded into the LQT by ablating a metallic Yb target or a pressed, annealed BaCl$_2$ target with a pulsed Nd:YAG laser. The assays are exposed to laser beams to Doppler cool a single isotope of either Ba$^+$ or Yb$^+$. These laser beams are roughly aligned along the axis of the LQT and retro-reflected at the exit viewport. For Ba$^+$ (Yb$^+$), a Doppler cooling beam at 493 nm (369 nm) and a repump beam at 650 nm (935 nm) are required. In the case of Ba$^+$, a magnetic field is applied to destabilize dark states, which result from the $\Lambda$ system [31]. Due to their Coulomb interaction with the laser-cooled species, other atoms and molecules in the trap are sympathetically cooled and the full assay reaches temperatures of (1–100) mK. Subsequently, the RF voltage is turned off within a fraction of an RF cycle and the electrodes are pulsed to DC high voltages (HV) with a 10 %–90 % rise time of $\approx 250$ ns. The HV is applied such that a two-stage electric field is established [8], which radially extracts the cold atoms and molecules from the LQT into the TOF-MS [32, 33]. This is accomplished by applying a slightly lower HV to the electrodes which are closer to the TOF-MS (1.2 kV) than to the ones that are farther (1.4 kV). The TOF drift tube has a length of 275 mm and its entrance is determined by a grounded skimmer with an aperture diameter of 5.6 mm. It contains two Einzel lenses which allow the focusing of the molecules. For the experiments presented here, a voltage of $\approx 900$ V ($\approx 500$ V) is applied to the front (rear) Einzel lens. The molecules are detected with a channel electron multiplier (CEM), which is shielded from the drift tube by a grounded stainless steel mesh.

## III. RESULTS

To precisely measure the attainable mass resolution, metallic Yb, having seven naturally abundant isotopes, is assayed in a first demonstration of LA-MS. The Yb target is ablated and $^{174}$Yb$^+$ ions are laser cooled (LA-MS[$^{174}$Yb$^+$]). The average assay size is $\sim 1000$ ions. Once laser cooled, we extract the assay into the TOF drift tube and record the TOF spectrum. The average of 20 spectra for the LA-MS[$^{174}$Yb$^+$] experiment is depicted in Fig. 2 (blue curve) and compared to the spectrum of the traditional TOF-MS without laser cooling (red curve). As the flight time $T$ and mass $m$ of a molecule are con-nected through $T \propto \sqrt{m}$, we use the dominant signal of $^{174}$Yb$^+$ corresponding to a flight time of $T_{174} \approx 9.84\,\mu$s and mass $m_{174} = 173.94$ Da for mass calibration [34].

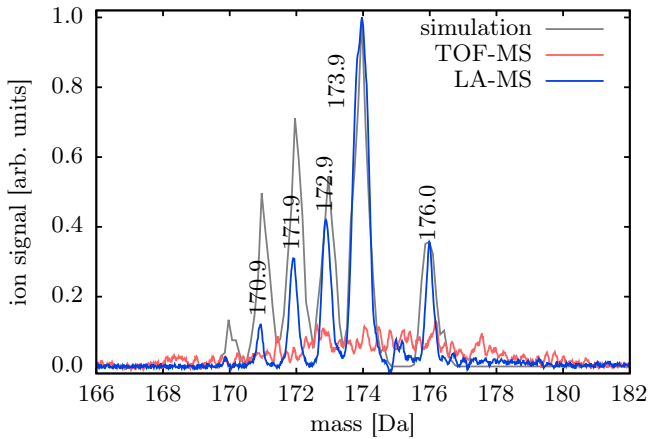

Figure 2. LA-MS[$^{174}$Yb$^+$] versus traditional TOF-MS for trapped Yb ablation products. Samples consisting of ≈ 1000 ions are loaded into the LQT and $^{174}$Yb$^+$ is laser cooled. The LA-MS curve represents the average of 20 spectra. The labels denote the mass of the ions of the corresponding peak and are determined by a fit of a Gaussian curve to the LA-MS spectrum. For comparison, a traditional TOF-MS curve, which is an average of 20 spectra taken without laser cooling, and a simulation of a LA-MS spectrum (see text) are shown.

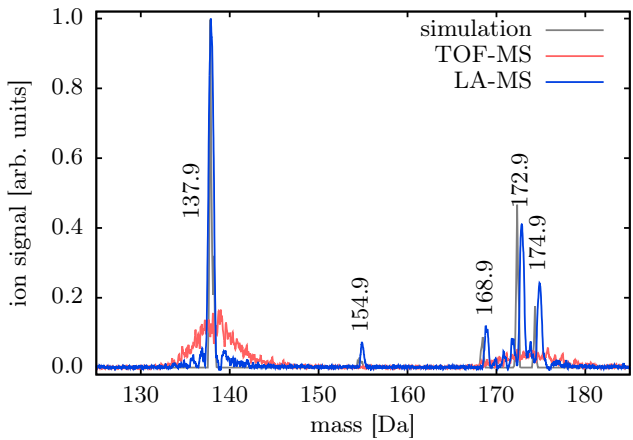

Figure 3. LA-MS[$^{138}$Ba$^+$] versus traditional TOF-MS for trapped BaCl$_2$ ablation products. Compared to Fig. 2, slightly larger assays are used for the LA-MS (of the order of few thousand ions) and $^{138}$Ba$^+$ is laser cooled. The LA-MS curve represents the average of 20 spectra. The labels denote the mass of the ions of the corresponding peak and are determined by a fit of a Gaussian curve to the LA-MS spectrum. For comparison, a traditional TOF-MS curve, which is an average of 20 spectra taken without laser cooling, and a simulation of a LA-MS spectrum (see text) are shown.

The peaks in the LA-MS[$^{174}$Yb$^+$] spectrum have a full-width-at-half-maximum of $\Delta m_{\text{LA-MS}} \approx 0.35$ Da. This corresponds to a mass resolution of $m/\Delta m \approx 500$ and allows isotopic resolution. All peaks can be explained by the naturally abundant isotopes of ytterbium: $^{168}$Yb (< 1 %), $^{170}$Yb (3 %), $^{171}$Yb (14 %), $^{172}$Yb (22 %), $^{173}$Yb (16 %), $^{174}$Yb (32 %), $^{176}$Yb (13 %) [35]. The peak in the traditional TOF-MS spectrum is a superposition of peaks belonging these Yb$^+$ isotopes. Hence, its width of ≈ 7 Da does not directly reflect the mass resolution of the traditional TOF-MS, but corresponds to a single-isotope width of $\Delta m_{\text{TOF-MS}} \approx 5$ Da and mass resolution of $m/\Delta m_{\text{TOF-MS}} \approx 35$. Compared to the traditional TOF-MS, LA-MS constitutes an improvement in mass resolution of more than an order of magnitude. Though the recorded signal intensities are also enhanced by more than an order of magnitude over traditional TOF-MS, the LA-MS signals are so large that they lead to partial saturation of the CEM. Thus, it is difficult to measure the true gain in detection sensitivity, however, simulation suggests an improvement for LA-MS of ∼ 30× over traditional TOF-MS with 300 K assays. Saturation of the detector also explains the broadening of the $^{174}$Yb$^+$ peak ($\Delta m_{174} \approx 0.5$ Da) and prevents reliable estimates of the abundances of different molecules in our assay; according to fluorescence images, we expect a higher abundance of $^{174}$Yb$^+$ than the peak heights suggest. As there is no naturally abundant $^{175}$Yb, the small peak at $m \approx 175$ Da could be due to the presence of $^{174}$Yb$^1$H$^+$ in our assays, or an electronics artifact resulting from the large, preceding signal, which is supported by the additional, slight undershooting before this peak.

Fig. 2 also shows the result of a molecular dynamics simulation [36] of LA-MS[$^{174}$Yb$^+$]. For this simulation, the electric field of the LQT is solved using a boundary-element method [37]. The simulation includes micromotion of the ions trapped in the LQT, mutual ion–ion Coulomb interaction, and the experimental, time-dependent voltages applied to the electrodes of the LQT. The curve is the average of 20 simulated spectra with an assay size of ≈ 500 Yb$^+$ ions with natural isotopic abundances. As the Einzel lenses are not included in the simulation and the experimental voltages are only known within 3 %, the simulated flight times show a deviation by ≈ 0.5 µs from the flight times observed in the experiment. Hence, we perform a separate mass calibration for the simulated data. The simulation agrees well with the experimental results and predicts a detection efficiency of 50 %. However, as the Einzel lenses are not considered in the simulation, this is likely a lower bound on the detection efficiency. For larger assays this number goes down slightly due to the axially elongated assay being clipped by both the skimmer and the aperture of the CEM.

In the second LA-MS demonstration, we analyze the ablation products of BaCl$_2$ to demonstrate the effectiveness of LA-MS for sympathetically cooled constituents of an assay over a broader mass range. Here, we laser cool $^{138}$Ba$^+$ (LA-MS[$^{138}$Ba$^+$]) and use an assay size of ∼ 1000 ions. The average of 20 spectra for the LA-MS[$^{138}$Ba$^+$] is shown in Fig. 3 (blue curve) and again compared to the traditional TOF-MS spectrum without any laser cooling

(red curve). The $^{138}\text{Ba}^+$ peak is used for mass calibration ($m_{138} = 137.9\,\text{Da}$) and leads to the same calibration as for the (experimental) LA-MS[$^{174}\text{Yb}^+$] spectrum.

The peaks in the LA-MS[$^{138}\text{Ba}^+$] spectrum at $m = 172.9\,\text{Da}$ and $m = 174.9\,\text{Da}$ can be explained by the presence of $^{138}\text{Ba}^{35}\text{Cl}^+$ and $^{138}\text{Ba}^{37}\text{Cl}^+$, respectively. The natural abundance of $^{138}\text{Ba}$ amounts to 72 % (other Ba isotopes are lighter and have at most 11 % natural abundance [35]) and chlorine naturally consists of $^{35}\text{Cl}$ (76 %) and $^{37}\text{Cl}$ (24 %). Hence, these $\text{BaCl}^+$ molecules represent by far the most abundant isotopic pairings and the signal intensities qualitatively match the natural abundances of the respective isotopes. The peak at $m = 154.9\,\text{Da}$ most likely represents $^{138}\text{Ba}^{16}\text{O}^1\text{H}$, see Ref. 30. The peak at $m = 168.9\,\text{Da}$ appears to be due to $^{138}\text{Ba}^{16}\text{O}^{12}\text{C}^1\text{H}_3^+$, a reaction product of barium and methanol, as methanol was used in the preparation of the $\text{BaCl}_2$ target. Additionally, the LA-MS[$^{138}\text{Ba}^+$] spectrum unveils the presence of the lighter $\text{Ba}^+$ iosotopes next to the $^{138}\text{Ba}^+$, $^{138}\text{Ba}^{35}\text{Cl}^+$, and $^{138}\text{Ba}^{37}\text{Cl}^+$ peaks. However, the peaks at $m \approx 139\,\text{Da}$ and $m \approx 176\,\text{Da}$ are, as already observed in the LA-MS[$^{174}\text{Yb}^+$] spectrum, either electronics artifacts or hydride ions.

The simulation of LA-MS[$^{138}\text{Ba}^+$] is performed analogously to the simulation of LA-MS[$^{174}\text{Yb}^+$]. Each individual simulation considers a total of $\sim 500$ ions comprising the above mentioned molecules and the curve in Fig. 3 is an average of 20 such simulated spectra. The deviation between experimental data and simulation is slightly larger than for LA-MS[$^{174}\text{Yb}^+$]. Presumably, this can be explained due to the time-dependent acceleration during the turn-on of the extraction high voltages, which can result in a (slight) departure of the mass calibration from the ideal $T \propto \sqrt{m}$ relation. As we know the extraction voltages only within 3 %, the magnitude of the observed deviation is not surprising.

For the data of Figs. 2 and 3, Einzel lenses are used to focus molecules with trajectories deviating from the axis of the drift tube onto the CEM, increasing the detection efficiency. These trajectories are longer than those of the on-axis molecules, leading to a broadening in the mass spectrum in traditional TOF-MS [30]. However, when using LA-MS we have observed that while the use of the Einzel lenses does improve the detection efficiency, it does not affect the mass resolution. Presumably, this is a result of the localization and cooling of the molecules before injection into the TOF-MS.

## IV. DISCUSSION

The observed improvement in mass resolution with cooling can be easily quantified for an ideal Wiley-McLaren TOF-MS, using the experimental parameters of the LA-MS[Yb$^+$] experiment, as shown in Fig. 4. Here, only the projection of the position and velocity onto the direction of extraction is considered. The mass resolution is deduced from the difference of the flight times of two ions with opposing initial conditions $(x, v) = \pm(\Delta x, \Delta v)$, while ion–ion repulsion is implemented using an estimated, averaged charge density of the ion cloud. As expected, the mass resolution increases for both a decrease of the width of the velocity $\Delta v$ and spatial distribution $\Delta x$, until the latter is limited by ion–ion repulsion within the trapping potential of the LQT (labelled Coulomb forbidden). Due to ion–ion repulsion, over confinement, $\Delta x \lesssim 100\,\mu\text{m}$ or an ion density of $\sim 10^8\,\text{cm}^{-3}$, can even lead to a slight decrease of mass resolution as the large of amount of potential energy between the ions is released during extraction. Points on this graph in the $\Delta x$–$\Delta v$ plane represent thermal distributions belonging to different cooling methods employed in mass spectrometry. This simple calculation agrees with the observed mass resolution for the traditional TOF-MS, buffer-gas cooling from previous studies using the TOF-MS in Ref. 30, and LA-MS.

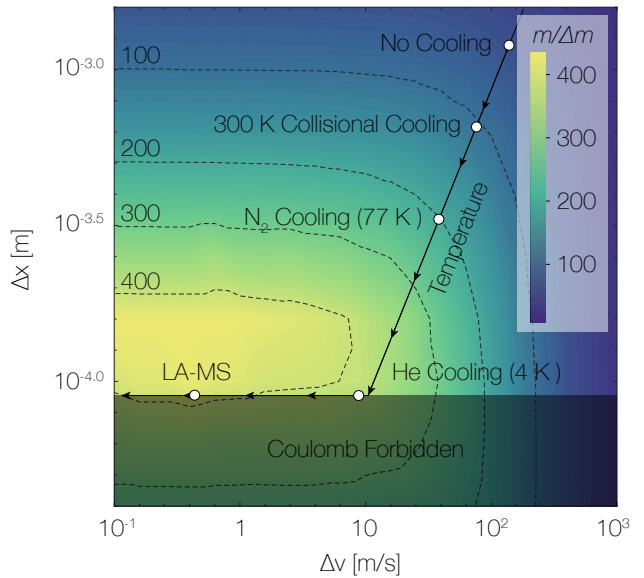

Figure 4. Mass resolution of an ideal Wiley-McLaren TOF [8] as function of the width of the velocity ($\Delta v$) and spatial distribution ($\Delta x$) of the input assay. The simulation assumes an input assay consisting of 1000 $^{174}\text{Yb}^+$ ions and LQT parameters as used in Fig. 2. The ion–ion Coulomb repulsion prohibits further squeezing of the assay's spatial distribution at increasingly lower temperatures (Coulomb forbidden region). Points correspond to distributions belonging to different cooling methods.

The mass range over which LA-MS is effective is set by the mass dependence of the ion trap pondermotive potential and the dynamics of the ion–ion sympathetic cooling in this potential. The pondermotive potential is inversely dependent on particle mass, leading to a practical limitation to the heaviest molecule that can be trapped and analyzed. In our experiment with Yb$^+$, we have verified that the trap depth can be changed within a factor of $\sim 10$. Therefore, we expect [38] that LA-MS[Yb$^+$] is ef-fective for masses up to $10^3$ Da. The lightest molecule that can be assayed is set by the dynamics of the ion–ion sympathetic cooling. Because of micromotion interruption it is increasingly difficult to sympathetically cool an ion as its mass becomes smaller than the laser cooled ion [39, 40]. Therefore, the mass range for LA-MS is limited between masses similar to the coolant ion—as low as 2 Da for LA-MS[Be$^+$] [41]—up to $\sim 10^3$ Da. This includes volatile organic compounds, the amino acids, explosive agents, and some peptides; modifications may extend the mass range to include heavier biomolecules such as nucleic acids [38, 42].

While the current TOF-MS implementation is basic and the mass resolution does not, by far, reach the resolution of state-of-the-art mass spectrometers (for example, > 100 000 for the Orbitrap [23]), LA-MS appears to be applicable to these mass spectrometers, where it could lead to an even higher mass resolution and/or improved detection limit. However, further work is necessary as phenomena like collective oscillations [43] in Fourier transform mass spectrometry, may preclude some types of mass spectrometers from realizing the full benefits of LA-MS. We emphasize that for LA-MS to be effective it is not necessary to cool the assay close to the Doppler cooling limit, nor produce ion Coulomb crystals [44]. Even laser detunings of several natural linewidths $\Gamma$ (e.g. $\Gamma_{Yb} \approx 2\pi \times 20$ MHz for Yb$^+$) from the optimal detuning for Doppler cooling do not affect the mass resolution of LA-MS, suggesting that LA-MS could be easy to implement in commercial mass spectrometers, especially, in comparison to cryogenic cooling.

In the presented form, LA-MS represents an easy-to-implement detection method in cold ion experiments that can be used complementarily to (and, perhaps, more robustly than) techniques based on analysing fluorescence images [38, 45–48] or mass spectrometry based on resonant excitation of the secular ion motion [26, 49, 50]. It is an ideal tool for spectroscopy of molecular ions and significantly outperforms previous implementations [51–54]. LA-MS also promises to be an ideal tool for the rapidly emerging field of hybrid atom–ion interactions [55–57]. It allows for large optical access required for these experiments, while providing isotopic identification of every trapped atomic/molecular ion. The current implementation of LA-MS will be used in ongoing studies of cold reactions of ionic atoms/molecules with neutral atoms [58–60], where it enables the distinction between different reaction products, and efforts towards cooling molecular ions in a hybrid atom–ion trap [61].

In summary, we have demonstrated a new method, LA-MS, to prepare assays of molecules for mass spectrometry, which outperforms conventional buffer-gas cooling. LA-MS improves the mass resolution of a single TOF-MS with comparatively short drift tube of 275 mm from $m/\Delta m \approx 35$ to $m/\Delta m = 500$ and enables isotopic resolution. LA-MS also increases the detection efficiency of the mass spectrometer by over an order of magnitude. Saturation effects of the CEM and its comparatively small aperture set the detection limit in our current setup. Using a micro-channel plate detector (MCP) with a larger active area, simulations suggest it is possible to detect every input ion with isotopic resolution.

## ACKNOWLEDGEMENT

We thank Peter Yu for assistance in the development of the LQT and TOF-MS electronics, and John Bollinger for guiding discussions. This work was supported by the ARO Grant No. W911NF-10-1-0505 and NSF Grant No. PHY-1005453.

[1] J. R. Yates III, Nat. Meth. **8**, 633 (2011).
[2] "Mass spectrometry: Limitless innovation in analytical science. market forecast: 2013–2017 (table of contents)," (2013), Strategic Directions International, Inc.
[3] S. Dresen, S. Kneisel, W. Weinmann, R. Zimmermann, and V. Auwärter, J. Mass Spectrom. **46**, 163 (2011).
[4] M. Hermanns-Clausen, S. Kneisel, B. Szabo, and V. Auwärter, Addiction **108**, 534 (2013).
[5] B. F. Cravatt, G. M. Simon, and J. R. Yates III, Nature **450**, 991 (2007).
[6] S. A. Hofstadler and K. A. Sannes-Lowery, Nat. Rev. Drug Discovery **5**, 585 (2006).
[7] H.-J. Kluge, Eur. J. Mass Spectrom. **16**, 269 (2010).
[8] W. C. Wiley and I. H. McLaren, Rev. Sci. Instrum. **26**, 1150 (1955).
[9] M. B. Comisarow and A. G. Marshall, Chem. Phys. Lett. **25**, 282 (1974).
[10] B. A. Mamyrin, V. I. Karataev, D. V. Shmikk, and V. A. Zagulin, Zh. Eksp. Teor. Fiz. **64**, 82 (1973).
[11] J. H. J. Dawson and M. Guilhaus, Rapid Commun. Mass Spectrom. **3**, 155 (1989).
[12] Q. Hu, R. J. Noll, H. Li, A. Makarov, M. Hardman, and R. Graham Cooks, J. Mass Spectrom. **40**, 430 (2005).
[13] M. Yamashita and J. B. Fenn, J. Phys. Chem. **88**, 4451 (1984).
[14] J. B. Fenn, M. Mann, C. K. Meng, S. F. Wong, and C. M. Whitehouse, Science **246**, 64 (1989).
[15] M. Karas, D. Bachmann, and F. Hillenkamp, Anal. Chem. **57**, 2935 (1985).
[16] K. Tanaka, H. Waki, Y. Ido, S. Akita, Y. Yoshida, T. Yoshida, and T. Matsuo, Rapid Commun. Mass Spectrom. **2**, 151 (1988).
[17] F. Hillenkamp, M. Karas, R. C. Beavis, and B. T. Chait, Anal. Chem. **63**, 1193A (1991).
[18] L. Schweikhard, S. Guan, and A. G. Marshall, Int. J. Mass Spectrom. Ion Processes **120**, 71 (1992).
[19] J. W. Hager, Rapid Commun. Mass Spectrom. **16**, 512 (2002).

[20] I. V. Chernushevich and B. A. Thomson, Anal. Chem. **76**, 1754 (2004).
[21] A. N. Krutchinsky, A. V. Loboda, V. L. Spicer, R. Dworschak, W. Ens, and K. G. Standing, Rapid Commun. Mass Spectrom. **12**, 508 (1998).
[22] A. V. Loboda, S. Ackloo, and I. V. Chernushevich, Rapid Commun. Mass Spectrom. **17**, 2508 (2003).
[23] A. Makarov, E. Denisov, A. Kholomeev, W. Balschun, O. Lange, K. Strupat, and S. Horning, Anal. Chem. **78**, 2113 (2006).
[24] D. Douglas and J. French, J. Am. Soc. Mass Spectrom. **3**, 398 (1992).
[25] D. J. Larson, J. C. Bergquist, J. J. Bollinger, W. M. Itano, and D. J. Wineland, Phys. Rev. Lett. **57**, 70 (1986).
[26] T. Baba and I. Waki, Jpn. J. Appl. Phys. **35**, L1134 (1996).
[27] M. Bussmann, U. Schramm, D. Habs, V. Kolhinen, and J. Szerypo, Int. J. Mass Spectrom. **251**, 179 (2006).
[28] R. Cazan, C. Geppert, W. Nörtershäuser, and R. Sa¡nchez, Hyperfine Interact. **196**, 177 (2010).
[29] Z. Andelkovic, R. Cazan, W. Nörtershäuser, S. Bharadia, D. M. Segal, R. C. Thompson, R. Jähren, J. Vollbrecht, V. Hannen, and M. Vogel, Phys. Rev. A **87**, 033423 (2013).
[30] S. J. Schowalter, K. Chen, W. G. Rellergert, S. T. Sullivan, and E. R. Hudson, Rev. Sci. Instrum. **83**, 043103 (2012).
[31] D. J. Berkeland and M. G. Boshier, Phys. Rev. A **65**, 033413 (2002).
[32] C. L. Jolliffe, "Mass spectrometer with radial ejection," (1996), patent, US 5576540.
[33] J. Franzen, "Method and device for orthogonal ion injection into a time-of-flight mass spectrometer," (1998), patent, US 5763878.
[34] G. Audi and A. Wapstra, Nucl. Phys. A **595**, 409 (1995).
[35] A. A. Sonzogni, Am. Inst. Phys. Conf. Proc. **769**, 574 (2005).
[36] T. Matthey, T. Cickovski, S. Hampton, A. Ko, Q. Ma, M. Nyerges, T. Raeder, T. Slabach, and J. A. Izaguirre, ACM Trans. Math. Softw. **30**, 237 (2004).
[37] C. Rosales Fernández, *depSolver v2.1*, Institute of High Performance Computing, Singapore (2008).
[38] S. Schiller and C. Lämmerzahl, Phys. Rev. A **68**, 053406 (2003).
[39] T. Baba and I. Waki, Appl. Phys. B **74**, 375 (2002).
[40] K. Chen, S. T. Sullivan, and E. R. Hudson, arXiv **1310.5190v1**, 1 (2013), 1310.5190.
[41] P. Blythe, B. Roth, U. Fröhlich, H. Wenz, and S. Schiller, Phys. Rev. Lett. **95**, 183002 (2005).
[42] D. Trypogeorgos and C. J. Foot, arXiv **1310.6294**, 1 (2013), 1310.6294.
[43] K. Jungmann, J. Hoffnagle, R. G. DeVoe, and R. G. Brewer, Phys. Rev. A **36**, 3451 (1987).
[44] G. Birkl, S. Kassner, and H. Walther, Nature **357**, 310 (1992).
[45] C. B. Zhang, D. Offenberg, B. Roth, M. A. Wilson, and S. Schiller, Phys. Rev. A **76**, 012719 (2007).
[46] P. F. Staanum, K. Højbjerre, P. S. Skyt, A. K. Hansen, and M. Drewsen, Nat. Phys. **6**, 271 (2010).
[47] S. Kahra, G. Leschhorn, M. Kowalewski, A. Schiffrin, E. Bothschafter, W. Fuß, R. de Vivie-Riedle, R. Ernstorfer, F. Krausz, R. Kienberger, and T. Schaetz, Nat. Phys. **8**, 238 (2012).
[48] Y.-P. Chang, K. Długołęcki, J. Küpper, D. Rösch, D. Wild, and S. Willitsch, Science **342**, 98 (2013).
[49] M. Drewsen, A. Mortensen, R. Martinussen, P. Staanum, and J. L. Sørensen, Phys. Rev. Lett. **93**, 243201 (2004).
[50] A. Ostendorf, C. B. Zhang, M. A. Wilson, D. Offenberg, B. Roth, and S. Schiller, Phys. Rev. Lett. **97**, 243005 (2006).
[51] K. Chen, S. J. Schowalter, S. Kotochigova, A. Petrov, W. G. Rellergert, S. T. Sullivan, and E. R. Hudson, Phys. Rev. A **83**, 030501 (2011).
[52] K.-K. Ni, H. Loh, M. Grau, K. C. Cossel, J. Ye, and E. A. Cornell, arXiv **1401.5423v1**, 1 (2014), 1401.5423.
[53] C. M. Seck, E. G. Hohenstein, C.-Y. Lien, P. R. Stollenwerk, and B. C. Odom, arXiv **1402.0123v1**, 1 (2014), 1402.0123.
[54] P. Puri, S. J. Schowalter, S. Kotochigova, and E. R. Hudson, "Action spectroscopy of $SrCl^+$ using an integrated ion trap time-of-flight mass spectrometer," to be published.
[55] A. Härter and J. Hecker Denschlag, Contemp. Phys. **55**, 33 (2014).
[56] S. Willitsch, arXiv **1401.1699v1**, 1 (2014), 1401.1699.
[57] C. Sias and M. Köhl, arXiv **1401.3188v1**, 1 (2014), 1401.3188.
[58] W. G. Rellergert, S. T. Sullivan, S. Kotochigova, A. Petrov, K. Chen, S. J. Schowalter, and E. R. Hudson, Phys. Rev. Lett. **107**, 243201 (2011).
[59] S. T. Sullivan, W. G. Rellergert, S. Kotochigova, and E. R. Hudson, Phys. Rev. Lett. **109**, 223002 (2012).
[60] S. T. Sullivan, W. G. Rellergert, S. Kotochigova, K. Chen, S. J. Schowalter, and E. R. Hudson, Phys. Chem. Chem. Phys. **13**, 18859 (2011).
[61] W. G. Rellergert, S. T. Sullivan, S. J. Schowalter, S. Kotochigova, K. Chen, and E. R. Hudson, Nature **495**, 490 (2013).